\documentclass[12pt,preprint]{aastex}

\slugcomment{To appear in: ApJ Lett.}

\shorttitle{Glitches from collision events}
\shortauthors{Huang \& Geng}

\begin{document}


\title{Anti-Glitch Induced by Collision of a Solid Body \\
       with the Magnetar 1E 2259+586}


\author{Y. F. Huang\altaffilmark{1} and J. J. Geng\altaffilmark{1}}
\affil{Department of Astronomy, Nanjing University, Nanjing 210093, China}
\email{hyf@nju.edu.cn}


\altaffiltext{1}{Key Laboratory of Modern Astronomy and Astrophysics (Nanjing University),
                 Ministry of Education, China.}


\begin{abstract}
Glitches have been frequently observed in neutron stars. Previously, these
glitches unexceptionally manifest as sudden spin-ups that can be explained
as due to impulsive transfer of angular momentum from the interior superfluid
component to the outer solid crust. Alternatively, they may
also be due to large-scale crust-cracking events. However, an unprecedented
anti-glitch was recently reported for the magnetar 1E 2259+586, which 
clearly exhibited a sudden spin-down, strongly challenging previous glitch
theories. Here we show that the anti-glitch can be well explained by the collision
of a small solid body with the magnetar. The intruder has a mass of about
$1.1 \times 10^{21}$ g. Its orbital angular momentum is assumed to be
antiparallel to that of the spinning magnetar, so that the sudden spin-down
can be naturally accounted for. The observed hard X-ray burst and decaying
softer X-ray emission associated with the anti-glitch can also be reasonably
explained. Our study indicates that a completely different type of glitches
as due to collisions between small bodies and neutron stars should exist and
may have already been observed previously. It also hints a new way for
studying the capture events by neutron stars: through accurate timing
observations of pulsars.
\end{abstract}





\section{Introduction}

Neutron stars are compact objects with typical mass 
$M_{\rm{ns}} \sim 1.4 {\rm M}_{\odot}$ and radius $R _{\rm{ns}}
\sim 10^6$ cm. They usually appear as radio pulsars,
with surface magnetic field $B_0 \sim 10^{11}
- 10^{12}$ G. There also exist a small amount of magnetars with
$B_0$ significantly larger than $4.4
\times 10^{13}$ G \citep{Thompson1995,Mereg2008,Olausen2013}.
Glitches have been observed in both normal
pulsars \citep{Wang2000,Espi2011,Yu2013} and magnetars
\citep{Kaspi2000,Kaspi2003,Dib2008,Living2010,Eichler2010,Gav2011}. Previously these glitches
unexceptionally manifest as sudden spin-ups that may be due to
impulsive transfer of angular momentum from the interior superfluid component
to the outer solid crust \citep{Aderson1975,Pines1985,Pizzo2011}, 
or caused by large-scale crust-cracking
events.  However, recently an unprecedented anti-glitch from the
magnetar 1E 2259+586 was reported \citep{Archibald2013}, which 
clearly exhibited a strange sudden spin-down.

Anti-glitches could be due to either an internal or an external
mechanism. For example, an impulsive angular momentum transfer
between regions of more slowly spinning superfluid and the crust
can produce the anti-glitch \citep{Thompson2000}. An external
model such as strong outflows \citep{Tong2013}, or a sudden
twisting of the magnetic field lines \citep{Lyutikov2013}, or
accretion of retrograde matter \citep{Katz2013, Ouyed2013} can
also cause the spin-down. However, most of these models involve
gradual deceleration processes. They cannot generate a sudden
spin-down and cannot account for the associated hard X-ray burst.
Also, they can hardly explain the extreme rarity of anti-glitches.

In this study, we propose a completely different external mechanism for the anti-glitch.
We suggest that the sudden spin-down could be due to the collision of a
solid body with the magnetar. Our model can reasonably explain the associated hard
X-ray burst and the decaying softer X-ray emission.

\section{Anti-glitch as observed}

The anomalous X-ray pulsar 1E 2259+586 is a magnetar ($B_0 = 5.9 \times 10^{13}$ G)
with a rotation period of
about 7s (rotation frequency $\nu \sim 0.143$ Hz) and frequency derivative
of $\dot{\nu} = -9.8 \times 10^{-15}$ Hz~s$^{-1}$, lying at a distance
$d = 4 \pm 0.8$ kpc \citep{Tian2010}. Historically, two major spin-up
glitches were observed in this AXP in 2002 \citep{Woods2004} and 2007 \citep{Icdem2012}.
In April 2012, the pulse time-of-arrival (TOA) of 1E~2259+586
experienced a strange anomaly that is most prominently characterized by a clear
sudden spin-down. The overall behavior can be described
by two possible timing scenarios \citep{Archibald2013}. In the first, there is an
instantaneous change in frequency by $\Delta \nu = -4.5(6) \times 10^{-8}$ Hz
on about 18 April. This spin-down glitch was then followed by a
spin-up glitch of amplitude $\Delta \nu = 3.6(7) \times 10^{-8}$ Hz
that occurred about 90 days later. In the second scenario, an anti-glitch of
$\Delta \nu = -9(1) \times 10^{-8}$ Hz occurred on 21 April. Then another
anti-glitch of amplitude $\Delta \nu = -6.8(8) \times 10^{-8}$ Hz
happened about 51 days later. Note that in this scenario,
the amplitudes of the two anti-glitches are comparable.

The two scenarios can fit the TOA data almost equally well (Archibald et al. 2013, 
however, see \citet{Hu2013} for a slightly different data analysis).
It seems somewhat surprising that two such
very different scenarios could be consistent with the same data. The difference
is actually due to the action of the persistent spin frequency derivative.
In the first scenario, the fitted derivative is
$\dot{\nu} \sim -3.7 \times 10^{-14}$ Hz~s$^{-1}$.
This enhanced spin-down episode lasts for a long period of about 90 days,
leading to an overmuch spin-down. So, it needs a normal spin-up glitch to
compensate for the excess. In the second scenario, the fitted derivative is
$\dot{\nu} \sim -2.3 \times 10^{-14}$ Hz~s$^{-1}$, and this spin-down episode lasts only for
about 51 days, which is clearly insufficient. So, it further needs
another spin-down anti-glitch to be in accord with the late TOA data.

For the above two scenarios, additional observational facts may help us to 
decide which one adhere to the truth better. Note that a hard
X-ray burst with a duration of about 36 ms was detected by Fermi/GBM
on 21 April, 2012 \citep{Foley2012}, just consistent with the epoch of the
preceding anti-glitch. The observed fluence of $\sim 6 \times 10^{-8}$ erg~cm$^{-2}$
in the 10 --- 1000 keV range corresponds to an energy release of
$E_{\rm xb} \sim 1.1 \times 10^{38}$ erg. An increase in the 2 -- 10 keV flux by a
factor of 2 was also observed to be closely related to the
anti-glitch \citep{Archibald2013}. It decayed continuously as a power-law function of
time in $\sim 260$ days. A simple integration then gives an extra energy
release of $E_{\rm x} \sim 2.1 \times 10^{41}$ erg during this
epoch. The flux increase was accompanied by a moderate
change in the pulse profile \citep{Archibald2013}. On the contrary, as for the succedent
glitch/anti-glitch event, no associated radiative or
profile changes were recorded. It strongly indicates that the
preceding event and the succedent event should be very different in nature. We thus
believe that the first scenario, i.e. an anti-glitch plus a normal
glitch, is more reasonable. We will carry out our study based on this description.

\section{Model}

\subsection{Small body - neutron star collision}

We propose that the sudden spin-down could be due to the collision of a small solid
body with the magnetar. The planetesimal has a mass of $m_{\rm pl}$. It headed for
the magnetar along a retrograde parabolic orbit, with a periastron distance of $p$ 
(see Fig. 1). When coming to the periastron, its velocity will
be $V_{\rm pl} = (2 G M_{\rm ns} / p)^{1/2}$, where $G$ is the gravitational constant.
The orbital angular momentum is $- m_{\rm pl} \cdot V_{\rm pl} \cdot p$.
We assume that the planetesimal was captured by the magnetar.
Conservation of angular momentum then gives
\begin{equation}
I_{\rm c} \cdot 2 \pi \nu - m_{\rm pl} V_{\rm pl} p = I_{\rm c} \cdot 2 \pi (\nu - \Delta \nu),
\label{eq:AM}
\end{equation}
where $I_{\rm c}$  is the moment of inertia of the
neutron star crust and all stellar components that are rigidly
coupled to it. Here we will first take $I_{\rm c} \sim 0.01 I_{\rm tot}
\sim 10^{43}$ g~cm$^2$ as a typical value
\citep{Pizzo2011, Hooker2013}, with $I_{\rm tot}$ being the moment
of inertia of the whole star. But in Section 5, we will give some
discussion on the other extremity that the superfluid in the core
strongly couples to the crust as a whole.
Equation (1) can be further simplified as
\begin{equation}
m_{\rm pl} \sqrt{2 G M_{\rm ns} p} = 2 \pi I_{\rm c} \cdot \Delta \nu.
\label{eq:simplify}
\end{equation}

The collision is a very complicate process. 
Tidal heating and Ohmic dissipation heating may happen. Part of the planetesimal 
will be evaporated, ionized and lost. The pulsar may even be temporarily 
quenched \citep{Cordes2008,Mottez2013a,Mottez2013b}. 
Here in our analysis, we would omit many of the subtle effects for simplicity.
Falling of the solid body onto the compact star can lead to the release of a binding
energy of $G M_{\rm ns} m_{\rm pl} / R_{\rm ns}$. The efficiency of transferring this
energy into {\em prompt} high temperature radiation (i.e., a burst) is very small for
a neutron star without a magnetic field.
However, the strong magnetic field of 1E 2259+586 can help to 
increase the efficiency significantly \citep{Colgate1981}. Also, although the impact process
may manifest as a series of falling-backs and re-expansions, the majority of
the binding energy should finally be deposited as thermal energy onto the star crust,
leading to an enhanced and much prolonged X-ray afterglow \citep{Harwit1973}.
This may correspond to the decaying X-ray emission associated with the
anti-glitch of 1E 2259+586. As mentioned before, the observed extra energy release
connected to the anti-glitch is
$E_{\rm xb} + E_{\rm x} \sim E_{\rm x}$.
Taking $E_{\rm x} \sim  G M_{\rm ns} m_{\rm pl} / R_{\rm ns}$, we can
get the required mass of the planetesimal as $m_{\rm pl} \sim 1.1 \times 10^{21}$ g.
Substituting $m_{\rm pl}$ into Equation (2), we further get the periastron distance
as $p = 1.7 \times 10^4$ cm. Of course, in this case, the solid body will
collide with the neutron star before coming to the periastron (see Fig. 1).
The off-axis distance of the impact point on the neutron star
surface (i.e. the impact parameter) is $b \sim 2.6 \times 10^5$ cm.

We now give a more detailed description of the collision. For simplicity,
we assume that the planetesimal is a homogeneous iron-nickel body with density
$\rho_{\rm pl} = 8$ g~cm$^{-3}$ \citep{Colgate1981}. Its radius is
then $r_{\rm pl} = 3.2 \times 10^6$ cm. Originally, the planetesimal is
a sphere. When approaching the neutron star, it will elongate 
due to strong tidal force. The maximum likely shear strength of the Fe-Ni body
is $S = 10^{10}$ dyn~cm$^{-2}$. So it will be broken up 
at a distance of $\sim 1.2 \times 10^{10}$ cm given by \citep{Colgate1981}
\begin{equation}
R_{\rm b} = (\rho_{\rm pl} r_{\rm pl}^2 M_{\rm ns} G / S)^{1/3} .
\label{eq:break_up}
\end{equation}
When the material finally pushes through the strong
magnetic field, it will be compressed to a thin sheet with the thickness of only a
few millimeters and density up to $10^6$ g~cm$^{-3}$. It also stretches significantly
in length. As a result, the time difference of arrival at the neutron star surface
becomes \citep{Colgate1981}
\begin{equation}
\Delta t_{\rm a}  =  \frac{2 r_{\rm pl}}{3} \cdot
       \left (\frac{2 G M_{\rm ns}}{R_{\rm b}} \right )^{-1/2} .
\label{eq:time}
\end{equation}

Taking $m_{\rm pl} = 1.1 \times 10^{21}$ g, a binding energy of $2.1 \times 10^{41}$
erg will be released during the fierce impact. 
Most of the energy will be reconverted to kinetic energy
due to plume development excited by the collision. However, even a
small portion of $\sim 5\times 10^{-4}$ being radiated in the 10 --- 1000
keV range will be enough to account for the observed hard X-ray
burst ($E_{\rm xb} \sim 1.1 \times 10^{38}$ erg) connected to the
anti-glitch. The calculated duration of $\Delta t_{\rm a} = 12$ ms
from Equation (4) is slightly less than that of the burst (36 ms). But 
it is acceptable since $\Delta t_{\rm a}$
may mainly correspond to the rising phase of the hard X-ray burst. 
Other portion of the binding energy will finally be
deposited as thermal energy onto the crust after complicated
processes. The diffusion of heat can lead to a power-law decay of
softer X-ray emission on a relatively long timescale
\citep{Lyub2002}, which may correspond to the power-law decay of
the 2 --- 10 keV flux after the anti-glitch.

\subsection{Origin of the small body}

Since a neutron star can retain the
planetary system during the violent supernova explosion that gives birth to
it \citep{Wols1994}, we speculate that there could be various
possibilities for the solid body. First, asteroids could be gravitationally
disturbed by other planets and be scattered toward the central
star \citep{Buren1981,Guill2011}. Second, like in our Solar System, there might also
exist circumstellar Oort-like clouds around the neutron star. Comets in these
regions can also fall toward the central star due to disturbance of nearby
stars \citep{Trem1986,Downs2013}. Third, in a system with multiple planets, the planets
may have chances to collide with each other and produce some clumps with a negative
angular momentum \citep{Katz1994,Ford2008}. Fourth, even if the neutron star escapes the
planetary system due to a large kick velocity,
it will take the runner $\sim$ 2400 years to pass through the planetary system and
the Oort-like clouds. During this period, the probability of capturing small bodies
should be considerable \citep{Zhang2000}.
Finally, a neutron star, due to its proper motion in space, may occasionally
encounter other stars that possess a comet cloud, and may experience an episode of
copious collisions \citep{Pineault1989,Shull1995}.

When modeling 1E 2259+586, we derived a relatively small periastron distance
of $p = 1.7 \times 10^4$ cm. But actually, the capture radius can be much larger. 
When a solid body passing
through the magnetosphere of a pulsar, strong Alfv\'en waves will be excited
which can carry away angular momentum very quickly. As a result, the material
falls onto the neutron star if \citep{Trem1986}
\begin{equation}
p  \leq  \left (\frac{9}{32 \pi^2 G c^2} \cdot \frac{B_0^4
   R_{\rm ns}^{12}}{r_{\rm pl}^2
   \rho_{\rm pl}^2 M_{\rm ns}} \right )^{1/9}.
\label{Alfven}
\end{equation}
For a magnetar, the capture distance can be $\sim 20 R_{\rm ns}$.
Also note that small bodies with $p$ up to $\sim 80
R_{\rm ns}$ (or possibly even larger) will be disrupted on their
first passage and then accreted on their second or subsequent
passages \citep{Trem1986, Livio1987}. Additionally, if the solid
body was of icy composition with $\rho_{\rm pl} \sim 1$
g~cm$^{-3}$, the allowed distance will further
increase by about 1.5 times.

Recently, the interaction between a relativistic pulsar wind and
the orbiting small body were studied in great detail by Mottez \& Heyvaerts (2011a, b).
It is found that Alfv\'en wings structures will be formed when the 
planet moves in the centrifugally driven wind. As a result, the
orbit will drift at a rate of \citep{Mottez2011b}
\begin{equation}
\left| \frac{{\rm d}a}{{\rm d}t} \right| \sim \frac{16 \pi r_{\rm pl}^2 R_{\rm ns}^4 B_0^2 \nu }
     {\mu_0 c m_{\rm pl} \sqrt{G M_{\rm ns} a^3}},
\label{drift}
\end{equation}
where $a$ is the semi-major axis and $\mu_0$ is the magnetic permeability
of vacuum.  Note that Equation (6) needs a correction when the eccentricity is not zero.
For a prograde orbit, $a$ increases and the orbit becomes more distant,
but for a retrograde orbit, $a$ decreases. The effect is very significant for a planetesimal
with a diameter $\leq 100$ km. For example, for a retrograde small body with $r_{\rm pl}=30$ km,
$m_{\rm pl} = 10^{21}$ g, and $a = 10^{12} $ cm, we have ${\rm d}a / {\rm d}t \sim  -2.8 \times
10^{9}$ cm/yr. A retrograde planetesimal thus could be captured in less than $\sim 1000$ years 
even if it is initially at a distance of $\sim 0.1$ AU. 
This effect can markedly increase the capture rate.

However, the exact event rate is very difficult to calculate due to many
uncertainties concerning the planetary system of pulsars. For example, a
preliminary estimate by \citet{Mitro1990} gives a wide range of one event
per 5,000 --- $3 \times 10^7$ years for a single neutron star,
depending on various assumptions of the capture radius, the relative
velocity at infinity, and the number density of small bodies.
In some special cases such as during comet showers, the event rate can
even be as high as $\sim$ 1 per year \citep{Trem1986,Zhang2000,Livio1987}.
For the whole Milky Way, \citet{Wasserman1994} argued that of order 0.1 --- 1
collisions may happen daily in the halo if the mass  
function extends continuously from brown dwarfs to asteroids.

\section{Explanation of the subsequent normal glitch}

For the subsequent normal spin-up glitch of amplitude
$\Delta \nu = 3.6(7) \times 10^{-8}$ Hz that occurred about 90 days later,
we suggest that it can be explained by usual glitch
mechanisms, such as the mechanism involving co-rotation of unpinned
vortices under weak drag forces \citep{Pizzo2011}. 
According to this mechanism, vortices in the superfluid star
core are only weakly pinned to the lattice of normal nuclear component. As the
neutron star slows down, vortices are continuously depinned and then
rapidly repinned. This dynamical creep can effectively shift the excess vorticity
outward on short timescales. The transferred core vorticity will be repinned in the
neutron star crust, where pinning force increases rapidly by orders of magnitude.
When the cumulated spin frequency lag exceeds the maximum
value ($\Delta \nu_{\rm max}$)  that can be endured by the crust, a sudden spin-up
glitch will happen. Major normal glitches were observed in 1E 2259+586 in 2002 and
2007, and this time in 2012. They show an obvious periodicity, which means the
typical interval between glitches is $\Delta t_{\rm gl} \sim 5$ yr. The maximum
frequency lag can then be calculated as $\Delta \nu_{\rm max} =
\Delta t_{\rm gl} \cdot |\dot{\nu}| = 1.5 \times 10^{-6}$ Hz \citep{Pizzo2011}.


In the crust, the maximum pinning force is gained when the density 
is $\rho_{m} \sim 0.2 \rho_0 $, with $\rho_{0}$ being the nuclear saturation density.
Define a dimensionless radius $x =
R / R_{\rm ns}$, then $\rho_{\rm m}$ corresponds to 
$ x_{\rm m} = 1 - 4 \rho_{\rm m} R_{\rm ns}^3 / (\pi M_{\rm ns}) \approx 0.97$.
The angular momentum stored
during $\Delta t_{\rm gl}$ and released at the glitch is
$\Delta L_{\rm gl} = I_{\rm \nu} (x_{\rm m}) \cdot \Delta \nu_{\rm max}$,
where $I_{\rm \nu} (x_{\rm m})$ is the effective moment of inertia.
Then, the amplitude of the glitch is \citep{Pizzo2011}
\begin{equation}
\Delta \nu = \frac{\Delta L_{\rm gl}}{I_{\rm tot} [1 - Q (1 - Y_{\rm gl})]},
\label{eq:amplitude}
\end{equation}
where $Q$ is the standard superfluid fraction, and 
$Y_{\rm gl}$ is a parameter that globally
describes the fraction of vorticity coupled to the normal crust on timescales
of the glitch rise time.

In Equation (\ref{eq:amplitude}), taking typical parameters 
as $Q = 0.95$ and $Y_{\rm gl} = 0.05$ \citep{Pizzo2011}, then we can get a
predicted glitch amplitude of $\Delta \nu = 3.1 \times 10^{-8}$ Hz for
1E 2259+586. It is in good agreement with the observed value of
$3.6(7) \times 10^{-8}$ Hz.

\section{Discussion}

The collisions between small bodies and neutron stars are basically possible.
The mechanism has been widely engaged to account for various transient
X/$\gamma$-ray events \citep{Colgate1981,Buren1981,Trem1986,
Livio1987,Pineault1989,Mitro1990,Katz1994,
Wasserman1994,Shull1995,Zhang2000,Cordes2008,Campana2011}.
In previous studies, attention is
ubiquitously paid on associated radiative activities. Here we suggest that
they potentially can also be diagnosed through accurate timing observations
of pulsars. This might be a more realistic way, 
since many pulsars are routinely monitored and the pulse TOA data are of
extremely high accuracy.

In our calculations, for the moment of inertia, we have
taken $I_{\rm c} \sim 0.01 I_{\rm tot}$. Although this is a reasonable
assumption for 1E 2259+586\citep{Kaspi2003,Icdem2012}, there are
also indications that in some pulsars, the superfluid in the core may be
strongly coupled to the crust on a very short timescale
\citep{Pines1985,Wang2000,Yu2013}. So, we now give some discussion on another
choice of $I_{\rm c} \sim I_{\rm tot}$. Since the mass of the small body is
determined from the observed X-ray fluence, $m_{\rm pl}$ in Equation (2) will
remain unchanged. We can then derive the periastron distance as
$p \approx 1.7 \times 10^8$ cm $\sim 170 R_{\rm ns}$. According to
the discussion in Section 3.2, the planetesimal may not be directly captured  
in this case, but it could be disrupted on its first passage and then accreted
on its second or subsequent passages. Especially, the magnetic thrust action
due to Alfv\'en wings \citep{Mottez2011a,Mottez2011b} may play a key role in 
the process, because the orbit drift rate could be as large 
as ${\rm d}a / {\rm d}t \sim -1.2 \times 10^{15}$ cm/yr according to Equation (6). 
So, an anti-glitch of similar amplitude will still happen.

The collisions between small bodies and neutron stars
can also lead to normal spin-up glitches, which is a completely different external
mechanism. In fact, considering the coplanarity of almost all planetary systems
observed so far, the chance of producing spin-up glitches should be much larger
than that for anti-glitches. We speculate that among
the several hundred normal glitches observed so far, some might actually be
collision events.

A basic feature of collision-induced glitches is that they are unlikely to show
any periodicity for a single neutron star. Also, they are more likely to happen
in young pulsars than in old pulsars, since the orbital motion of small bodies
might be more instable soon after the supernova explosion. Note that a
collision-induced glitch can be either radiatively active or silent. In
Equation (2), if we take $p = 40 R_{\rm ns}$, then a mass of
$m_{\rm pl} = 2.3 \times 10^{19}$ g will be enough to produce a glitch with the
amplitude similar to that of the anti-glitch observed in 1E 2259+586. In this
case, since $m_{\rm pl}$ is lower by a factor of $\sim$ 50, it is expected
that the associated X-ray burst will be very weak and hard to detect.

Finally, it is interesting to note that observational evidence 
for asteroids at a close distance to PSR B1931+24 was recently 
reported \citep{Mottez2013a,Mottez2013b}.  
For PSR J0738$-$4042, evidence of an asteroid interacting with the pulsar 
was also declared very recently \citep{Brook2013}. 
In the future, more observations would be available and should be helpful
for probing such collision events.

\acknowledgments

We appreciate many helpful comments and suggestions from an anonymous referee. 
We thank T. Lu, Q. H. Peng, Z. G. Dai, X. Y. Wang, F. Y. Wang and Ming Xu
for discussions, and Cong Cong for reading the manuscript.
This work was supported by the National Basic Research Program of China 
with Grant No. 2014CB845800, and by the National Natural Science
Foundation of China with Grant No. 11033002.

\begin{figure}[tbp]
\begin{center}
\includegraphics[width=70mm,angle=0,clip]{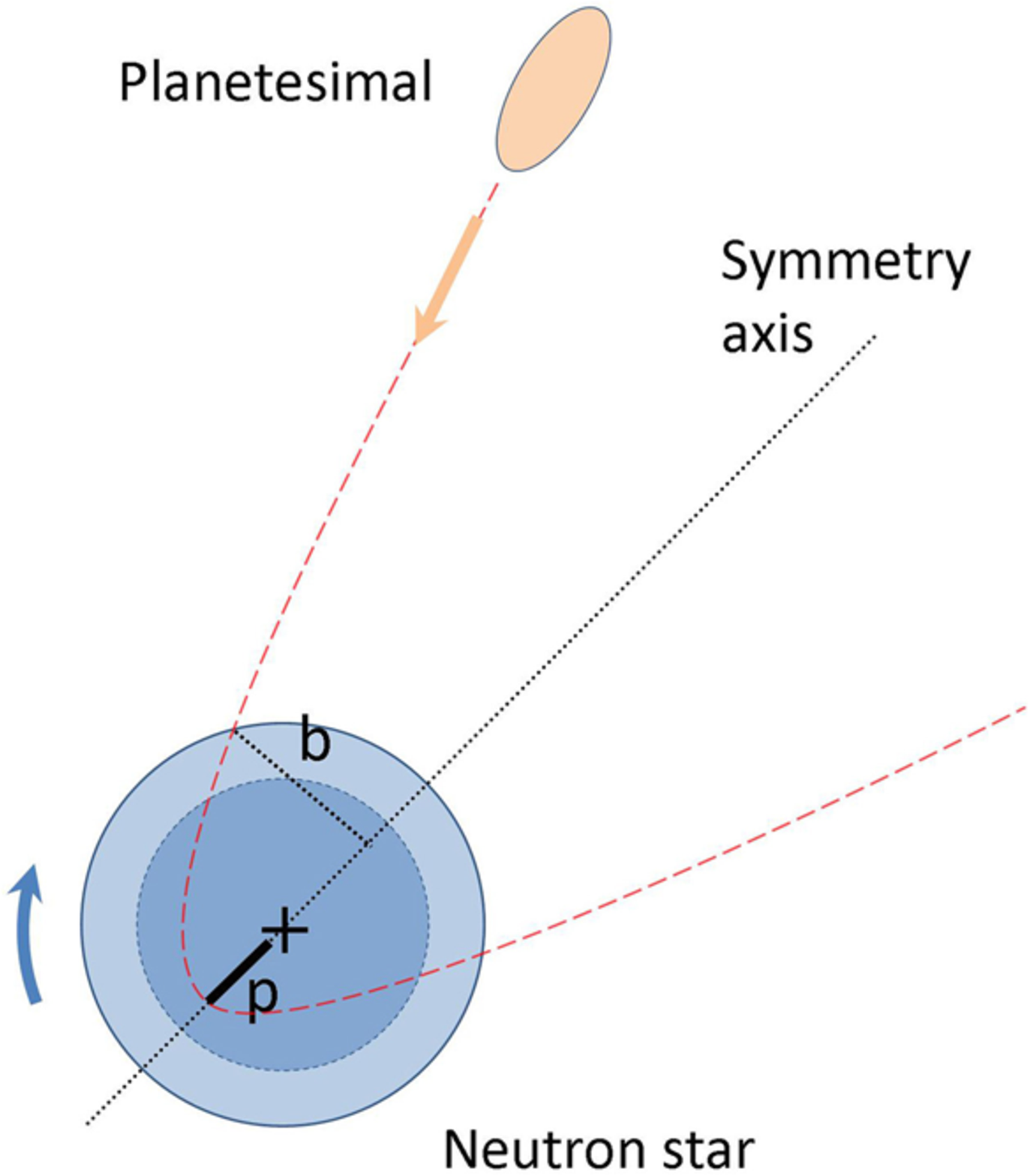}
\end{center}
\caption{Schematic illustration of the collision process. The deep blue circle is
the core of the neutron star, and the light blue ring represents the crust
(not to scale). The planetesimal heads for the neutron star along a retrograde 
parabolic orbit (red dashed curve) with a periastron distance of $p$. 
For 1E 2259+586, $p$ is derived as $\sim 1.7 \times 10^4$ cm, and
the impact parameter at the star surface is $b \sim 2.6 \times 10^5$ cm.
}
\label{fig:Cartoon}
\end{figure}








\end{document}